\documentclass[runningheads]{llncs}
\usepackage{algorithmic}

\usepackage{amsmath}

\usepackage{graphicx}

\usepackage{multirow}

\usepackage{xurl}
\usepackage{subcaption}

\usepackage[table,xcdraw]{xcolor}
\usepackage[many]{tcolorbox}

\usepackage{fontawesome}

\usepackage{hyperref}

 \usepackage{todonotes}
\usepackage{balance}
 \usepackage{flushend}
\definecolor{main}{HTML}{CFCFCF}  
\definecolor{sub}{HTML}{CFCFCF}   
\usepackage[framemethod=tikz]{mdframed}
\usepackage{etoolbox}

\definecolor{boxbg}{gray}{0.92}      
\definecolor{boxborder}{gray}{0.55}  

\newmdenv[
  skipabove=0.5\baselineskip,
  skipbelow=0.7\baselineskip,
  linewidth=0.5pt,
  linecolor=boxborder,
  backgroundcolor=boxbg,
  roundcorner=6pt,
  innerleftmargin=10pt,
  innerrightmargin=10pt,
  innertopmargin=8pt,
  innerbottommargin=8pt
]{boxC}

\newcommand{\boxCTitle}[1]{%
  \noindent\textbf{#1}\par\vspace{4pt}%
}

\newcounter{keyTakeAwaysCounter}

\newcounter{keyRQAnswerCounter}

\newcounter{RQCounter}

\newcounter{keyLimitationsCounter}

\AtBeginDocument{%
  }

\newcommand{\blue}{\textit{\textcolor{blue}{blue }}}
\newcommand{\orange}{\textit{\textcolor{orange}{orange }}}

\begin{document}

\title{Analyzing the Evolution of Structural Communities within Microservice Architecture}

\author{Alexander Bakhtin\inst{1}\orcidID{0000-0003-3513-7253}\and \\ Matteo Esposito\inst{1,2}\orcidID{0000-0002-8451-3668}\and 
 Valentina Lenarduzzi\inst{2,1}\orcidID{0000-0003-0511-5133}\and \\Davide Taibi\inst{2,1}\orcidID{0000-0002-3210-3990}}
\authorrunning{Bakhtin et al.}

\institute{ University of Oulu, Oulu, Finland \\  \email{{firstname.lastname}@oulu.fi},
\and University of Southern Denmark, Vejle, Denmark \\ \email{\{lastname\}@imada.sdu.dk}}


\titlerunning{Analyzing the Evolution of Structural Communities within MSA}

\maketitle

\begin{abstract}
In recent years, the detection of anti-patterns in microservice architecture has gained traction, particularly to identify instances of Microservice Architectural Degradation. In such tasks, the microservice architecture is often modeled as a network of microservice dependencies.

Temporal community detection methods have been proposed to analyze the community structure of networks that evolve in time.
We performed temporal community detection within the microservice architecture of six releases of the \textit{train-ticket} benchmark and analyzed the composition of the discovered communities and their activities over time. We observed a stable architecture with a clear separation of services into two communities, which we could identify with two business processes performed by the system.


\end{abstract}

\section{Introduction}
\label{sec:intro}
Over the past decade, the adoption of Microservice Architecture (MSA) has led to the identification of various architectural patterns and anti-patterns to prevent Microservice Architectural Degradation \cite{al2022using}.  Specific anti-patterns include Wrong cuts, when different responsibilities and business processes are incorrectly divided among the services, and Knot services, when services are too coupled to each other and communicate with too many other services \cite{cerny2023catalog}. Detecting such anti-patterns often involves modeling the microservice system as a network of connected services~\cite{bakhtin2025network,bakhtin2025hublike}. Moreover, in the context of architectural degradation, it is the \textit{evolution} of such networks that is of interest \cite{bakhtin2025ccp}.

In the field of network science, the study of \textbf{\textit{temporal networks} (TNs)} has gained traction as a powerful way to model and analyze networks that evolve \cite{holme2012temporal,bakhtin2025ccp}. 
Specifically in MSA, Bakhtin et al. have applied temporal network methods to networks of microservice architecture \cite{bakhtin2025ccp} and developer collaboration \cite{bakhtin2024temporal}.
In particular, temporal community detection allows to identify communities that persist or re-emerge through time \cite{gauvin2014detecting}.
In MSA networks, community detection can help us identify groups of services that communicate a lot and assess whether such groupings correspond to business processes \cite{gaidels2020service}, or if their dependencies are the result of improper division of responsibilities or unoptimized behavior \cite{genfer2026using,khodabandeh2024network}.

Therefore, leveraging the promising results obtained previously \cite{bakhtin2025ccp,bakhtin2024temporal}, in this work, we analyze the temporal community structure \cite{gauvin2014detecting} of several releases of \textit{train-ticket} OSS benchmark.
Our contributions include the following:
\textbf{(i)} we performed temporal community detection in the reconstructed MSA;
\textbf{(ii)} we analyzed the temporal activity of the discovered communities and noted stable community structure;
\textbf{(iii)} we observed the composition of microservice communities and how they reflect the business processes of the system.

Due to space constraints, background is provided in the Online Appendix \footnote{\url{https://doi.org/10.5281/zenodo.19253698}}.



\section{Related Work}
\label{sec:back}

Community detection among microservices within MSA is an unexplored direction. The only works that we could identify are due to Khodabandeh et al. \cite{khodabandeh2024network}, analyzing the community structure of microservice clusters in the Alibaba production environment, and Gaidels and Kirikova \cite{gaidels2020service}, who analyzed an MSA network with several community detection algorithms and identified the context of each community.
Crucially, \textit{temporal} community detection has not been leveraged for the assessment of microservice architectural evolution, while Bakhtin et al. \cite{bakhtin2025ccp} only applied temporal centrality for this purpose.
Abufouda and Abukwaik \cite{abufouda2017using} argued for properly adopting TN methods in Empirical Software Engineering. However, their review focused only on applications of TNs for developer collaboration networks. Bakhtin et al. \cite{bakhtin2024temporal} have leveraged temporal community detection to analyze the structure and trends of TN of developer collaboration on microservices in an OSS benchmark, determining the groups of collaborating developers at each release.
To our knowledge, this is the first attempt to apply temporal community detection to reconstructed microservice architecture networks in the context of Microservice Architectural Degradation.

\section{Empirical Study Design}
\label{sec:method}
The \textbf{goal} of our study is \textit{to \textbf{analyze} how the \textbf{temporal community structure} of microservices \textbf{evolves over time} and on what basis the microservices \textbf{communities are formed}.}
Therefore, we defined the following \textbf{Research Questions} (\textbf{RQs}): 

\textit{\textbf{RQ$_1$}: Do the structure and activity of the microservice communities evolve?
}

Since microservice architecture strives for \emph{high cohesion, low coupling} design of the system, the presence of strongly coupled groups of microservices in the MSA network
could indicate improper decomposition of the system into microservices or the presence of certain anti-patterns. 
Moreover, temporal variations in the community structure indicate that microservices change their dependencies from one release to the next, suggesting the instability of the architecture. Observing the trends in community activity and structure over time thus allows us to judge the stability of the architecture.
Since the approach of Gauvin et al. \cite{gauvin2014detecting} provides us with the communities' activity profiles over the analyzed time-span, we can observe if communities have near constant activity in the analyzed releases or if their activity varies a lot. We can leverage standard deviation for this analysis, and thus formulate the following null and alternative hypotheses:
\textit{\textbf{(H$_{01}$)} the activity values of each community stay within a 3-sigma interval around the mean;
\textbf{(H$_{11}$)} at least one community activity exceeds a 3-sigma interval around the mean.}

While the activity patterns of the discovered communities show the stability of the overall system, membership of particular services in different communities could serve as an indicator of Wrong cuts or Knot-like behavior:

\textit{\textbf{RQ$_2$}: Do microservices form communities based on business processes?
}

Once we have determined the composition of each community, we can analyze whether these communities consist of microservices performing related business processes. For example, in the case of the \textit{train-ticket} benchmark \cite{zhou2018benchmarking}, we could assume the services for booking a ticket, reserving a seat, and checking the timetable would be in the same community, since they would all be invoked when purchasing a ticket. Moreover, since the method by Gauvin et al. \cite{gauvin2014detecting} provides overlapping communities, i.e., the same service can belong to several communities, we can check if there are microservices that simultaneously belong to different communities focused on different business processes. If this is the case, it could hint at an improper service decomposition and responsibility allocation within the microservice system.
We thus formulate the following hypotheses:
\textit{\textbf{(H$_{02}$)} all microservices belong to exactly one community;
\textbf{(H$_{12}$)} at least one microservice belongs to several communities.}

\textbf{Data Collection.}
We have previously compiled a dataset of OSS MSS projects suitable for network analysis by reconstructing their architectures the Code2DFD tool \cite{schneider2023automatic,bakhtin2025network}.
In a follow-up study \cite{bakhtin2025ccp},
we determined the OSS MSS projects whose architecture evolved throughout development.
We discovered that the \textit{train-ticket}\footnote{\url{https://github.com/FudanSELab/train-ticket/}} microservice benchmark project is the only available and reconstructed system that shows significant structural evolution.
We thus leveraged the networks of all the train-ticket releases reconstructed in our previous work \cite{bakhtin2025network,bakhtin2025ccp} as the TN for our analysis.




\textbf{Data Analysis.}
We performed temporal community detection in the TN of \emph{train-ticket}, applying the method by Gauvin et al. \cite{gauvin2014detecting} with our previously developed implementation \cite{bakhtin2024temporal}.

To determine the best number of communities to fit, we leveraged the core consistency metric \cite{bro2003core} similar to \cite{bakhtin2024temporal,gauvin2014detecting}.
According to the original authors, the core consistency values between the maximum of 100 and 90 are considered a good fit \cite{gauvin2014detecting,bro2003core}.
For each choice of the number of communities from 2 to 6, we reconstructed the community structure 20 times with random initialization of PARAFAC decomposition, measured core consistency,
and computed their mean and standard deviation to determine the most stable results.
In our case, the only instance in which we obtain values above 90 is for the fit of \textbf{two communities} to the TN of \textbf{releases \emph{v0.0.1}-\emph{v0.2.0}}. This choice has an average core consistency of 99.05 and a standard deviation of 0.40.
We used this community detection result for our analysis. All the used scripts and data of all the tested values are available in the Online Appendix \footnote{\url{https://doi.org/10.5281/zenodo.19253698}}

We obtained the temporal activities of the two communities, as well as the membership strengths of microservices within these communities, based on incoming and outgoing connections.
We scaled all three outputs to the interval $[0.0;1.0]$ by dividing by the maximum value in each case.

\section{Results}
\label{sec:results}
\begin{figure}
    \centering
    \includegraphics[width=\linewidth]{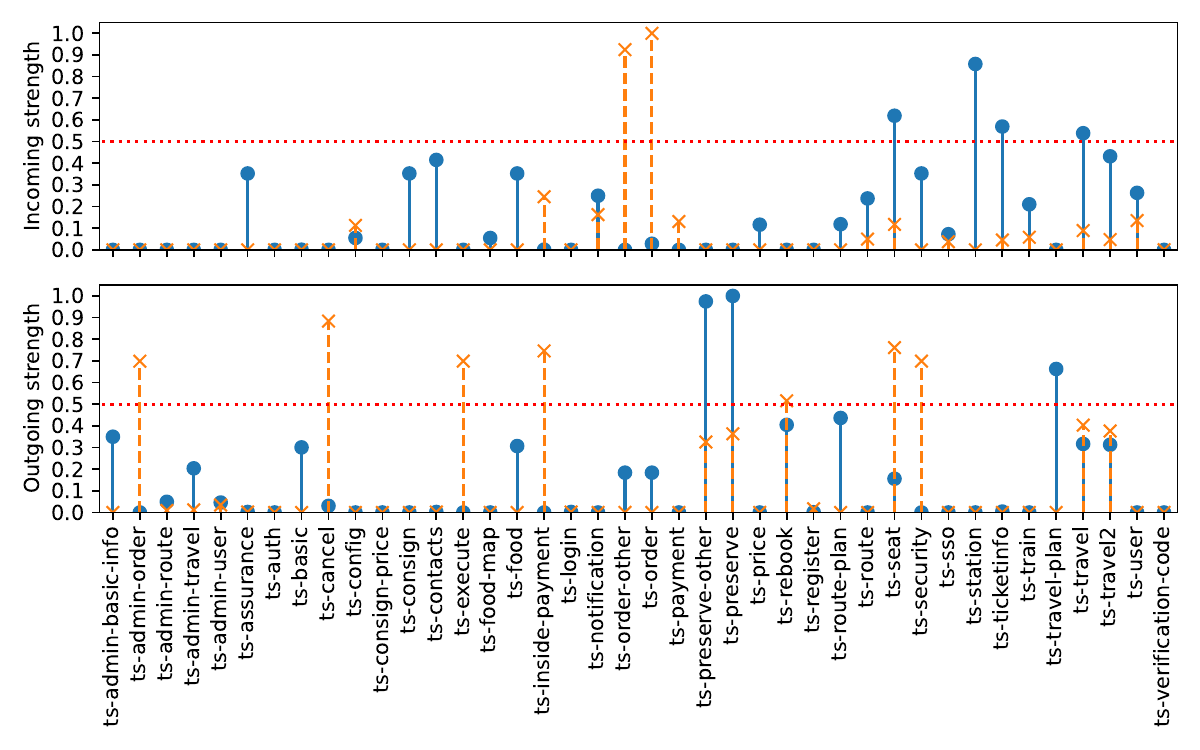}
    \caption{Incoming and outgoing community membership, normalized by the maximum value. The horizontal lines show the membership cutoff threshold of 0.5.}
    \label{fig:communities}
\end{figure}

In this section, we report the results of temporal community detection in the TN of \emph{train-ticket} architecture and answer the \textbf{RQs}.
The two detected communities are denoted consistently as \blue round, solid stems and \orange crossed, dashed stems in Figure \ref{fig:communities}.

Looking at the \textbf{temporal activity} of the communities (\textbf{RQ$_1$}, see Figure 1 in the Online Appendix), we observe that both \blue and \orange communities have \textbf{high activity}, consistently above $0.8$ in all the releases. The activity on the interval \emph{v0.0.2} - \emph{v0.1.0} is constant, indicating \textbf{identical, unchanging architecture}, as noted by Bakhtin et al. \cite{bakhtin2025ccp}. On this interval, \blue community \textbf{appears to be dominating} the \orange community, since the former increased its activity while the latter decreased compared to release \emph{v0.0.1}.

To answer \textbf{RQ$_1$}, we computed the mean and variance of the temporal activities of each community.
The exact temporal activity values of the \blue community are 0.99, 1.0, 1.0, 1.0, 1.0, 0.97, providing a mean of 0.993 and standard deviation of 0.012.
Thus, the 3-sigma interval is $[0.956;1.000]$.
For the \orange community, the values are 0.84, 0.82, 0.82, 0.82, 0.82, 0.81 with the mean 0.822, standard deviation 0.0098, and the 3-sigma interval $[0.793;0.851]$. Thus, for both communities, we \textbf{fail to reject the null hypotheses H$_{01}$}, since all temporal activity values stay within the respective 3-sigma intervals.

\textit{\textbf{RQ$_1$ answer}: We observe \textbf{stable community structure} across the releases of train-ticket, with all activity values within 3-sigma interval around the mean.}

    
    
    
    

Considering the \textbf{composition of the temporal communities} (\textbf{RQ$_2$}),
we determined the membership of microservices in the communities by setting a neutral threshold of 0.5 on the computed membership strength (Figure \ref{fig:communities}).

Observing the \blue community, we notice that it corresponds to \textit{preserve}, \textit{preserve-other}, and \textit{travel-plan} services making calls to services like \textit{seat}, \textit{station}, \textit{ticketinfo}, and \textit{travel2} services. The aim of the \textit{preserve} service is to preserve the ordered ticket data in the database, thus this community seems to correspond to the process of saving the different aspects of the order information, like stations mentioned in the ticket or the reserved seat, from respective services. 

Furthermore, the \orange community consists of several services like \textit{admin-order}, \textit{cancel}, \textit{rebook}, and \textit{seat} sending calls to \textit{order} and \textit{order-other} services. The \textit{order} services stores the information on the made orders, thus this community seems to correspond to the business process of modification of the orders, such as rebooking or cancellation.


Concerning our hypothesis, we have to \textbf{reject the null hypothesis} in favor of the alternative hypothesis H$_{12}$ since the \textit{seat} service is a member of both communities, \blue as a callee and \orange as a caller.


\textit{\textbf{RQ$_2$ answer:} Microservices participate in different communities associated with different business processes.}
 

\section{Discussion}
\label{sec:discussion}
In this section, we discuss the results of our \textbf{RQs}, their implications for practitioners and researchers, as well as address the threats to validity.

\textbf{Stability of the temporal community reconstruction.}
Overall, we have observed a \textbf{stable architecture} of the \emph{train-ticket} project, as indicated by the stable, almost \textbf{constant temporal activity} of the communities (\textbf{RQ$_1$}) and \textbf{clear separation of the microservices} into communities (\textbf{RQ$_2$}).
However, the fact that we could only fit a community structure
to the TN of releases \emph{v0.0.1}-\emph{v0.2.0}, excluding the latest release \emph{v1.0.0}, could itself indicate problems with the architecture stability due to sudden changes in the MSA network.
The change between releases \emph{v0.2.0} and \emph{v1.0.0} includes the refactoring of food-related services, which could have resulted in \textbf{noticeable changes in microservice dependencies} in that region of the MSA, and since the whole MSA consists of ``only" 42 services, this could have had a \textbf{significant effect on the performance} of the community detection algorithm.
We believe we would see more robust results if we considered an industrial-scale system with a long history of releases, since in an Agile environment, releases occur frequently and incrementally.
On the other hand, temporal activity was constant between the releases \textit{v0.0.2} and \textit{0.1.0}, which is the interval which we previously identified as having no modifications in the architecture \cite{bakhtin2025ccp}. Thus, the \textbf{leveraged method is robust} to constant data and produces stable, non-fluctuating results when given a temporal network with identical snapshots.

\textbf{Membership strength paints a nuanced picture.}
While we resorted to setting a threshold on the membership strength to interpret microservice communities and test our hypothesis, we also noticed that many services have non-zero membership strength below the $0.5$ threshold (Figure \ref{fig:communities}), and such observations could still provide value to architects and developers. 
Thus, the raw membership strength values can \textbf{enable more fine-grained interpretation}. 
For example, \textit{rebook} and \textit{route-plan} services have membership strength in the \blue community close to $0.5$. 
Moreover, services belong to several communities, suggesting that they \textbf{participate in several business processes}. This could indicate an inappropriate division of responsibilities between the services. Furthermore, some services have both the incoming and outgoing membership strength in the same community
, like \textit{food}, \textit{travel}, and \textit{travel2} services in the \blue community. 
This could indicate instances of \textbf{Wrong cuts, Knot-like, or Hub-like behaviour \cite{bakhtin2025hublike}}, since these services appear to be both called by and to call many other services.
We believe that performing this kind of analysis on a large industrial system will enable more detailed observation of communities and their identification with the business process. 
Moreover, with more communities detected overall, the membership of a microservice in several communities would be easier to connect to bad practices or anti-patterns, with many overlapping communities potentially indicating higher severity.

\textbf{Implications for researchers and practitioners.} The proposed temporal community analysis can provide valuable \textbf{insights to practitioners} who wish to analyze how their MSA is evolving. It could highlight \textbf{positive trends}, e.g., architecture remaining stable, or \textbf{negative trends}, e.g., services not related by a business process or function forming a community or a community significantly increasing its activity.
\textbf{For researchers, future directions} could include considering more closely what kinds of insights could be extracted from such an approach, 
in particular when analyzing a system that contains more communities that show sudden changes in activity,
and how they relate to anti-patterns such as Hub-like or Knot services or Wrong cuts \cite{bakhtin2025hublike,cerny2023catalog}. Moreover, the applicability of \textbf{other temporal community detection approaches} with different parameters, constraints, and outputs
could be considered.

\textbf{Threats to validity.}
We discuss the threats to the validity following the guidelines by Wohlin et al.~\cite{wohlin2024experimentation}.
\textbf{\textit{Construct validity:}}
we leveraged the dataset of several versions of the \textit{train-ticket} benchmark as reconstructed in our previous work \cite{bakhtin2025network,bakhtin2025ccp} with the Code2DFD tool \cite{schneider2023automatic} as well as our own implementation of the community detection methods \cite{bakhtin2024temporal}. These tools have their limitations as research prototypes.
\textbf{\textit{Internal validity:}}  
we had to rely on \textit{train-ticket} as the only MSA that is reconstructed for only seven releases with some architectural variability. However, even this variability is limited, with several releases having an identical architectural structure. 
Moreover, since the system contains only 42 microservices with many dependencies among them, we could only identify two communities, where a lot of services belong with a small, but non-zero strength. 
\textbf{\textit{External validity:}}
the \textit{train-ticket} benchmark project is one of the biggest OSS MSA projects, and the only one that we could analyze in this work. However, it still does not represent an industrial-scale, in-production microservice system. We were thus careful not to overgeneralize our observations and conclusions , aiming instead to showcase the leveraged community detection algorithm.
\textbf{\textit{Conclusion validity:} }
we are not closely familiar with all the services, functionalities, and business processes of \textit{train-ticket}, so the mapping between the communities and business processes is subject to interpretation bias.

\section{Conclusion}
\label{sec:conclusion}
In this work, we performed temporal community detection within a temporal MSA network of six releases of the \textit{train-ticket} benchmark project
and identified two communities with near-constant temporal activity.
Several services have incoming and outgoing memberships to several communities, which we identified with distinct business processes.
For \textbf{practitioners}, the proposed methodology could highlight positive and negative trends in MSA evolution, such as architecture remaining stable or services not related by a business process forming a community. \textbf{For researchers}, this work opens up future directions for studying how MSA community structure and service membership interplay with bad practices and anti-patterns. Furthermore, different community detection algorithms could be analyzed and compared. 

\noindent\textbf{Acknowledgment.}
This work has been funded by the Research Council of Finland (grants n. 359861 and 349488 - MuFAno) and Business Finland (6GSoft).


\bibliographystyle{splncs04}
\bibliography{bibliography}

\end{document}